# George Augustus Linhart – as a "widely unknown" thermodynamicist

E. B. Starikov[1,2]

[1]*Institute for Materials Science and Max Bergmann Center of Biomaterials, Dresden University of Technology, D-01062 Dresden, Germany,*
*E-Mail: starikow@tfp.uni-karlsruhe.de*

[2]*Department of Physical Chemistry, Chalmers University of Technology, SE-412 96 Gothenburg, Sweden*

**Abstract**

The name of George Augustus Linhart is in fact "widely unknown". In effect, he was a Viennese-born USA-American physicist-chemist, partially associated with the Gilbert Newton Lewis' school of thermodynamics at the University of California in Berkeley. As a lone small boy, he had arrived (from Austria via Hamburg) at New York in 1896, but was officially USA-naturalized only in 1912. He was able to pick up English in the streets of New York and Philadelphia, when occasionally working as a waiter and/or as a tailor – just to somehow survive. But, nonetheless, he could successfully graduate a high school in about one year – and then went to the universities for his further education. After obtaining his BS from the University of Pennsylvania, he could manage getting both MA and then PhD from the Yale University, Kent Chemical Laboratory. George Augustus Linhart was afterwards definitely able to successfully work out the true foundations of thermodynamics and could thus outdistance many famous thermodynamicists of his time and even the later ones. Linhart's view of the Second Law of Thermodynamics was and is extremely fruitful. The interconnection of Linhart's ideas with those of Gilbert Newton Lewis, as well as with the modern standpoints are discussed here in detail.

## Introduction

The name of George Augustus Linhart (May, 3, 1885 near Vienna, Austria - August, 14, 1951 in Los Angeles, CA) is "widely unknown" ... In effect, he was a Viennese-born USA-American physicist-chemist, partially associated with the Gilbert Newton Lewis' school of chemical thermodynamics at the University of California in Berkeley.

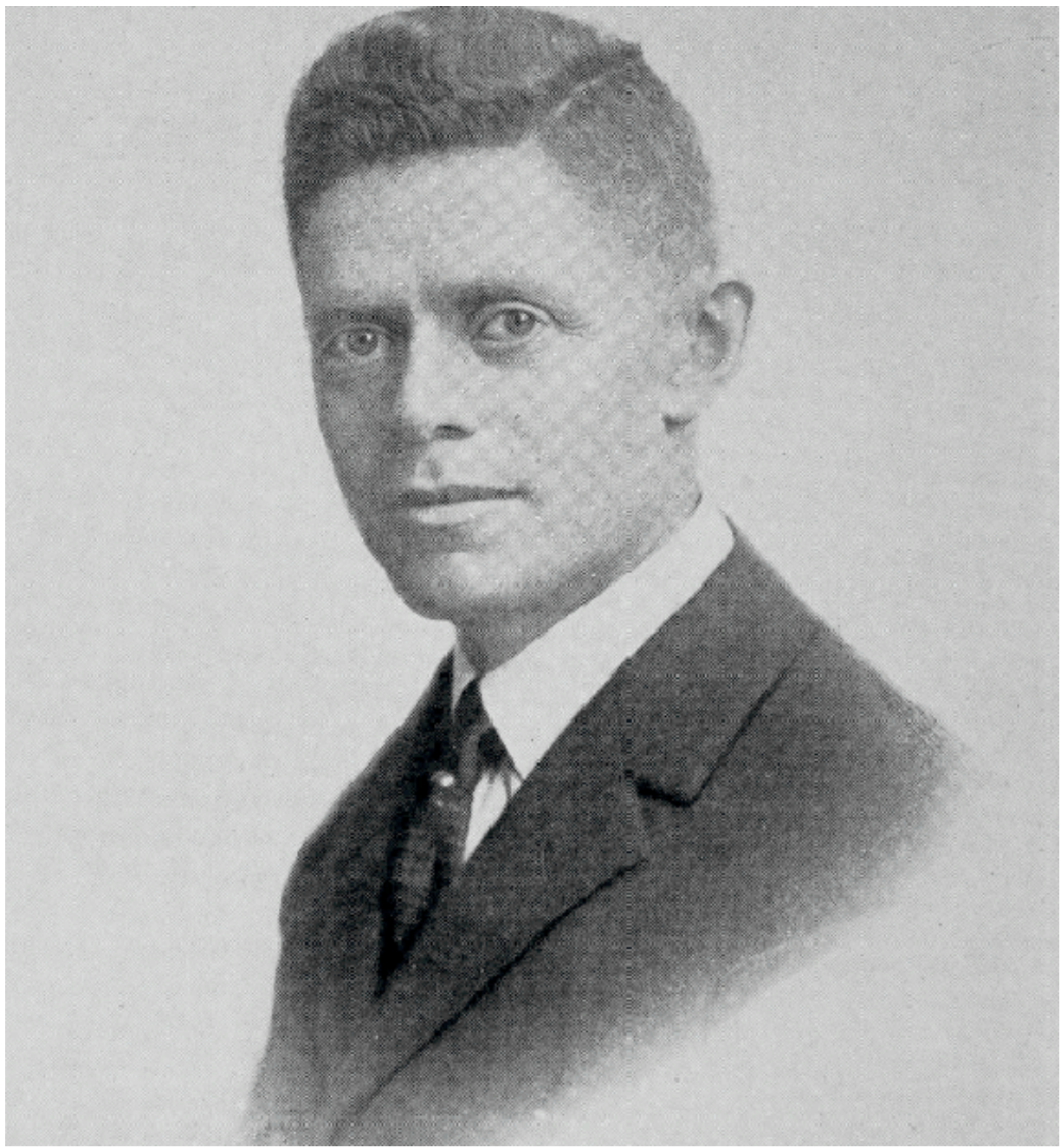

**George Augustus Linhart, in Riverside, California, around 1923**

After attentively reading G. A. Linhart's unpublished diaries ("Out of the Melting Pot" (1923) as well as "Amid the Stars" (1950)) and thoroughly digging all the possible USA archives, I have finally managed to reasonably reconstruct his extremely intensive, but, to my mind, still

at the utmost curvilinear and in fact very unlucky CV, which can roughly be sketched as follows:

As a lone small boy, he had arrived (from Austria via Hamburg) at New York in 1896, but was officially USA-naturalized only in 1912 … He was then married (in the twenties-thirties of the XX-th century), but, apparently, for some rather short time, and, as a result, had no children. As his life had started to come to an end, he left all his worldly possessions to different USA universites and scientific research organizations, just to endow fellowships for young scientists …

*Education*

He was able to pick up English in the streets of New York and Philadelphia, occasionally working as a waiter and as a tailor – just to somehow survive … But still could successfully graduate a high school in about one year – and then went to the universities for his further education …

BS from the University of Pennsylvania, 1909.
MA and PhD from the Yale University, Kent Chemical Laboratory (supervisor: Frank Austin Gooch): 1909-1913.

*Working*

University of Washington, Seattle:
Teaching Instructor in German and Chemistry, 1913-1914.

Simmons College, Boston:
Teaching Instructor, 1915.

University of California in Berkeley (he started working there in the lab of Gilbert Lewis):
Assistant, Chemistry, 1915-16;
Teaching Fellow, Chemistry, 1916-17;
Assistant, Chemistry, 1917-18;
Drafted for the World War I, 1918;
Assistant, Biochemistry, 1919 (Feb-May);
Instructor, Soil Chemistry and Bacteriology, 1919-20;
Research Associate, Soil Chemistry and Bacteriology, 1920 (May-July).

Eureka Junior College, Eureka, CA
Teacher, 1920-1921

Riverside Junior College, Riverside, CA
Teacher, 1921-1948

G. A. Linhart had successfully published about 20 papers (mostly on inorganic chemistry, as well as some treatises concerning his view on thermodynamics) in such journals as "American Journal of Science", PNAS, JACS, "Journal of Physical Chemistry" etc. But the most of his work and bright ideas is nevertheless contained in a number of unpublished preprints, to our sincere regret …

**The bright ideas of Georg(e) A. Linhart**

Although several formulations of the 2$^{nd}$ Law of Thermodynamics (2LT) different from each other are known, there still remains some kind of interpretational reticence. Specifically, everybody knows, on the one hand, that the 2LT law forbids the *perpetuum mobile* and this is empirically correct. On the other hand, the 2LT ought to predict that virtually everything in the Universe is perpetually running down, which is in apparent contradiction with a lot of well-known and observable natural phenomena characterized by not only disorganization and decay, but also self-organization and growth. Still, we know of a wealth of natural and technical processes, which are inherently irreversible, like the famous Humpty-Dumpty who "sat on a wall" and then "had a great fall". So, how could it be possible to bring all these facts under one and the same roof ?

Let us take a closer look at the essence of 2LT. First of all, we immediately see that the classical thermodynamics, which is the origin of this law, is applicable to equilibrium states only, where all the parameters of the system under study stop any changing. Any process in the classical thermodynamics *must* undergo a sequence of equilibrium states. The time during which such processes last is of no interest at all: they might take either five seconds or five hundred years to proceed, the main point is that everything happens in or in the nearest proximity to thermodynamic equilibrium. The 2LT states that such processes may sometimes be irreversible, so that there is absolutely no way for any spontaneous return to the initial state starting from the final one. The quantitative measure of processes reversibility is entropy: in isolated systems (isolated = no energy and/or matter exchange with surrounding) the latter either remains the same (reversibility) or increases (irreversibility).

Moreover, all the classical theoretical thermodynamics was originally derived for the cyclic equilibrium processes, where initial and final states coincide. Already the earliest attempts by Clausius and Kelvin to apply the 2LT to generic non-cyclic processes have immediately led to the speculations about the "*heat death of the Universe*" and, consequently, hot debates which are more or less ongoing even nowadays. The clou here is that all the physical laws at the microscopic level should always work same way even if the time course would change its direction. With this in mind, it is necessary to ensure that in truly macroscopic systems, where lots of microscopic particles come together, processes unidirectional in time become possible. The first huge advance in solving this fundamental problem was made by Boltzmann, who guessed (without any rigorous inference) that entropy should be nothing else but a logarithm of the number of all the microstates corresponding to one and the same observable macrostate. To this end, one might say that events like complete and perfect re-assembling of the broken Humpty-Dumpty are in principle possible but essentially unlikely, for there is a plenty of choices to destroy the poor Humpty-Dumpty, but relatively few ways of bringing him back to his authentic "initial state".

Meanwhile, there is much more to the story. First of all, a big confusion persists about the mathematical derivation of the 2LT-savvy expression for the entropy starting from the time-symmetrical microscopic physical laws (see, for example, [1]). The most radical standpoints even claim that the very notion of "irreversibility" has several different meanings (*time-asymmetry* is different from just *irrecoverability*) and one should not confound them as done usually (see, for example, [2, 3]). Independently of this, the line of work pioneered by Onsager, Prigogine, de Groot and Mazur is trying to redefine the conventional equilibrium notions of entropy, temperature etc. for the distinctly non-equilibrium situations, and a considerable progress was achieved on this way (see, for example, [4-6]). However, the main assumption of the latter works is that although everything in the Nature rarely reaches a perfect equilibrium, there ought to be some small areas allowing the conventional equilibrium description. To this end, it can be shown that the very notion of "thermodynamic equilibrium" should be re-conceived as a kind of fuzzy set describing continuous "degrees of equilibrium" (instead of the conventional crisp "equilibrium-nonequilibrium" binary picture) if we would like to bridge the conceptual gap between the Boltzmann and Gibbs thermodynamics [1,7]. Finally, the very notion of entropy causes serious debates as well (see, for example, [8] and the references therein).

To sum up, the nature of time and the interrelationship between time and entropy still remain a mystery. Here, the chief problem is "not to explain why the entropy of the universe will be higher tomorrow than it is today but to explain why the entropy was lower yesterday and even lower the day before that. We can trace this logic all the way back to the beginning of time in our observable universe. Ultimately, time asymmetry is a question for cosmology to answer" [9].

But is this really the case ? Have we already exhausted all the resources to tackle this problem upon Earth ? Surely not ! We may well refrain from Immediately Going to Heaven (or to the Hell, well, in accordance with the sums of our personal sins). Not only the gist, but even the mathematical details of this encouraging answer were given in the forgotten and/or unpublished works by George Augustus Linhart – approximately at the same time as the famous notion of "Time's Arrow" was coined by Sir Arthur Stanley Eddington in his well-known book "*Nature of the Physical World*" [10]. Doing justice and giving credit to that fundamental work by G. A. Linhart is the main purpose of the present communication.

The CV of G. A. Linhart was extremely dramatic (cf. the Introduction section above). He had graduated from Kent Laboratory of Chemistry in Yale and defended his PhD there as well, but all his professional life long, after several postdoc years in the University of Washington in Seattle and in Berkeley, he was all the rest of his life long working as a teacher in two countryside junior colleges in California: Eureka and Riverside. Bearing this in mind, the appallingly bitter introduction to his preprint "*The Relation Between Chronodynamic Entropy and Time*" [11] could be fully understandable:

*"Eddington in his delightful book on "The Nature of the Physical Universe" wonders, why entropy, so intimately associated with time, should be expressed quantitatively in terms of temperature instead of time. The present writer wondered about this too, for three consecutive years, during which time it was his good fortune to be in the very midst of multitudinous entropy calculations under the direction and guidance of G. N. Lewis. His first attempt at expressing his wonderment in mathematical form appeared in a rather humble periodical of approximately zero circulation. No wonder Eddington makes no reference to it. It was the Eureka Junior College Journal of Science, Arts and Crafts (1921)."*

I have thoroughly attempted fetching the above-mentioned work by G. A. Linhart, but really in vain – it seems to be completely lost for us. Fortunately, his preprint [11], which has been made available to me by courtesy of the colleagues in the Riverside Junior College, gives a full and detailed account of his bright ideas.

G. A. Linhart was considering general non-cyclic processes unidirectional in time and starts out with two ideas: *progress* and *hindrance*, which should underlie ***any*** process under study. What is the essence of these both ? G. A. Linhart answered:

*"By progress is meant any unidirectional phenomenon in nature, such as the growth of a plant or an animal, and by hindrance – the contesting and ultimate limitation of every step of the progress. In other words, progress is organized effort in a unidirectional motion, and hindrance is not so much the rendering of energy unavailable for that motion, as it is the disorganization of the effort to move; it acts as a sort of stumbling block in every walk of life. It is this property of matter which the writer wishes to measure quantitatively in relation to time, and which he designates as chronodynamic entropy."*

Hence, for us, the modern readers, G. A. Linhart insisted on the *dialectical* viewing of any physical process, with its initial and final stages being in general different from each other. Indeed, there always ought to be some *driving force* due to the transition of energy from one state to another, according to the 1$^{st}$ Law of Thermodynamics (1LT), which ensures and entails the *progress*. On the other hand, nothing all over the world would happen without the omnipresent *hindrance,* whose name is entropy, which is but nothing else than the genuine 2LT statement. The true dialectics of these both lies not only in the "universal competition between energy and entropy" [12,13], but also in their mutual compensation (cf., for example, [14-18] and references therein). Without intervention of the entropic *hindrance* no process would ever reach its final state, because the *progress* would then last forever. Thus, in fact, Linhart conceives entropy as just a kind of "Mephisto", whom J. W. von Goethe characterizes as *"ein Teil von jener Kraft, die stets das Böse will und stets das Gute schafft" ("Part of that Power which would the Evil ever do, and ever does the Good"* [transl. by G. M. Priest]). To this end, entropy represents solely *one integral part* of the "Yin-Yang"-tandem of thermodynamics [19], that is, the dialectic "energy-entropy" or "1LT-2LT" dyad: the entropy should not be considered separately from the energy and absolutized, as done conventionally. Furthermore, because any process is a result of the above-mentioned "unity and struggle of opposites", we may never invoke any guarantee that such processes would perfectly come to their final states as could be envisaged by theories. In fact, already the latter state of affairs should introduce the intrinsic fundamental indeterminacy in the sense of statistical causation of the processes [20]: in other words, any unidirectional process ought to require involvement of an essential probabilistic element, even without the conventional application to the atomistic build-up of the matter.

In accordance with all the above, G. A. Linhart's mathematical proposal is seemingly very simple: let us take the conventional definition of thermodynamic entropy, *q/T*, where *q* stands for the energy/heat spent during the process, *T* denotes the temperature, and *substitute the temperature by the time* in this ratio. This is why, the result of such a substitution was dubbed by G. A. Linhart "*chrono*dynamic entropy", instead of the initial "*thermo*dynamic entropy".

First of all, it is important to discuss the physical and mathematical lawfulness of such a substitution. Interestingly, an analogous trick has recently been independently and tacitly employed by Grisha Perelman to transform the "entropy-like" functional in his award-winning proof of the famous Poincaré conjecture, as a part of the full geometrization conjecture [21].

For the above purpose it would be suitable to use the foundational work by Caratheodory (In effect, very similar mathematical results on the foundations of thermodynamics were obtained by Gyula Farkas years before Caratheordory, but the Farkas' work remained unnoticed, probably because of its extraordinary terseness [22]). Although, strictly speaking, the original Caratheodory's inferences are only applicable to "simple" systems pertaining to thermodynamically equilibrium situations – and notwithstanding the proven intrinsically local character of Caratheodory's theorem [23, 24] – it was shown [25, 26] that the latter could be reasonably extended to define entropy, temperature etc. for rather generic nonequilibrium and irreversible cases. (Meanwhile, the works [25, 26] seem to be forgotten, like those by G. A. Linhart, in spite of their immense significance for the foundations of thermodynamics.) Meanwhile, the Lieb-Yngvason foundational work [27] is of little help for our present task, because it is blatantly absolutizing the notion of entropy.

To this end, if we assume the existence of the internal energy and the validity of the 1LT for generic processes, then the existence of entropy (and thus the generic validity of the 2LT) will follow as a direct consequence of the integrability of the 1LT differential form, to the effect that in the neighborhood of a thermodynamic state other states exist which are not accessible by reversible and adiabatic processes. Then, the entropy can be defined as

$$\partial Q = \lambda dF, \quad dF = \frac{\partial Q}{\lambda}, \tag{1}$$

where $\partial Q$ stands for the infinitesimal amount of heat, or anyway change in energy (and it is then an inexact differential), while $\lambda$ and $F$ are some non-zero functions of the variables of state (so that, the differential $dF$ is exact). The usual way of reasoning is to identify $F$ as entropy $S$ and the integrating factor $\lambda$ as an absolute temperature being the universal function of the empirical temperature only. But from the mathematics of the Pfaffian forms expressing the 1LT it is well known that, if there exists one integrating factor, then there is a wealth of them. Therefore, we may define a plenty of the "universal functions of empirical temperature" and then nothing stands in our way to identify *time* as one such, taking into account that, like temperature, time is an intensive variable, independent of system's dimension. To sum up, the foundations of thermodynamics do not forbid to have and employ a number of different entropies in connection with the multitude of the integrating factors, thus underlying the anthropomorphism of entropy [8, 28] and showing us, how to correctly handle the latter. Thus, the "bold tricks" of George and Grigori appear to be fully valid and applicable.

Now, we have approached the next important question: What is the sense and the use of the temperature-to-time substitution in the expression of entropy?

To answer this question, G. A. Linhart considers his theory based on the example of generic processes of growth, but underlines that such an approach is in fact exceedingly general. He starts out from the idea that the rate of spending the energy to promote the growth process ought to be proportional to the parameter measuring the progress of the process in question:

$$\frac{dE}{dt} = RG, \quad Q = dE = RGdt, \tag{2}$$

where $E$ is energy, $G$ stands for the mass of the growing body at some time $t$, $R$ – the proportionality coefficient between the mass and the energy (in general – between the

"driving force" and the "measurable progress"). Then, $Q$ is the analogue of the thermodynamical amount of heat, which is necessary to define the chronodynamic entropy, $Q/t$, that is, the degree of the process hindrance. If we cast the latter as $RdS$, we immediately get using Eq 2:

$$dS = \frac{Gdt}{t}. \tag{3}$$

On the other hand, owing to the dialectic interplay between the progress and hindrance, we may consider the ratio $G/G_i$, with $G_i$ being the hypothetical maximum mass achievable by the growing body, as a *probability* that the latter will arrive at the mass $G$ at some time $t$. Accordingly, the probability of the disjoint event, i. e., that the growing body will *not* arrive at this value of mass, is $1 - G/G_i$. With this in mind, G. A. Linhart proceeds to revealing the functional relationship between the probabilities thus defined and the chronodynamic entropy. His inference is based upon considering the dialectic nature of the "progress-hindrance" dyad. He states:

*"The question might be asked: "What is the degree of hindrance in any natural process before it occurs?" Obviously, there can be no hindrance to anything that does not exist. But at the instance of inception of the process hindrance sets in and continues to increase until ultimately it checks nearly all progress, and reduces to a minimum the chance of any further advance. At this juncture the outcome of the process is said to have approximately attained its maximum. It is clear then, that the increase in mass is in the same direction as the increase in hindrance, or entropy ..."*

To this end, G. A. Linhart mathematically arrives at the following expression for the ratio of the progress and hindrance infinitesimal increments:

$$\frac{dG}{dS} = K\left(\frac{G_i - G}{G_i}\right) = K\left(1 - \frac{G}{G_i}\right), \tag{4}$$

where the proportionality factor $K$ is the efficiency constant of the process (G. A. Linhart's reasoning here is as follows: "... *the smaller the value of K, the greater the hindrance ... and the slimmer the chance for the growing individual to survive.*"). The ratio $\left(\frac{G_i - G}{G_i}\right)$ is on the one hand, as G. A. Linhart puts it: "*the capacity of growth, which is unity at inception and statistically zero at completion*" – but on the other hand, it is the probability that the progress will *not* be achieved.

Hereafter, G. A. Linhart just employs the trivial straightforward mathematics. Indeed, combining Eqs 3 and 4, the rate of growth progress can be expressed as

$$\frac{dG}{dt} = \frac{K}{G_i}\left(\frac{G_i - G}{t}\right)G, \tag{5}$$

which on integration can be recast as:

$$\log \frac{G}{G_i - G} = K \log t + \log k, \tag{6}$$

or:

$$G = G_i \frac{kt^K}{1 + kt^K} = G_i \frac{x^K}{1 + x^K}; \quad x = \frac{t}{t_s}, \tag{7}$$

where $t_s$ is the *time scale* when measuring the time during the process under study, so that $x$ is the relative time.

The integrated form of Eq 4 is then:

$$S = \frac{2.3 G_i}{K} \log \frac{G_i}{G_i - G}, \tag{8}$$

which is nothing else, but the famous Boltzmann expression for the entropy, when we take into account the probabilistic interpretation of the ratio under the logarithm sign, whereas the final aim of the G. A. Linhart's theory, the expression of the chronodynamic entropy vs. time can be cast as follows, by combining Eqs 3 and 7 and integrating the result:

$$S = \frac{2.3 G_i}{K} \log(1 + x^K). \tag{9}$$

The immense significance of the simple and straightforward inference embodied in Eqs 2 – 9 consists in that G. A. Linhart has succeeded to mathematically derive the Boltzmann's expression for entropy, starting from the general dialectic ("Yin-Yang") principle. This is definitely a great advance in comparison with some attempts to derive the same formula starting from a number of purely mathematical axioms [29, 30]. The full correspondence of Eqs 8 and 9 to the Boltzmann's ingenious guess completely justifies somewhat artificial- and haphazard-looking linear approximation employed by G. A. Linhart in Eq 4.

Of considerable interest is Eq 6, which was not discussed by Linhart, but enables a unique interpretation of the nature of time. Specifically, we recast this equation in the following way, bearing in mind Eq 7:

$$x^K = \frac{G}{G_i - G} = \frac{P}{1 - P}, \quad P = \frac{G}{G_i}. \tag{10}$$

Eq 10 reveals that the relative time $x$, power the efficiency constant $K$, is nothing else but the *odds in favor* of the progress in the general growth process. The probability that the growing body will achieve the required mass, $P$, is hence defined *via* Dutch book argument, so that the G. A. Linhart's theory is firmly based upon the Bayesian epistemology. A huge advance here is achieved by the fact that *both* probabilities *and* the odds are measurable in experiment, so that, we do not need any additional theoretical or computer modeling, unlike in the conventional statistical mechanics. Advocating the above approach, G. A. Linhart has definitely outdistanced his time.

The interpretation of time as the "odds in favor of the progress" may sound extremely strange to us, yet it ought to contain a profound philosophic sense. Specifically, there is a definite parallel to the concept of time delivered by the ancient Chinese book "I-Ching", "The Book of Change" (see, for example, [31]). In fact, the I-Ching concept of the time, **si/shi/ji** (in korean/chinese/japanese), tells us, on the one hand, that "the time is something like the life force, current or pulse of a given set of circumstances", thus constituting an eternally present, all-pervasive and decisive element. Still, in the real life, there are nevertheless certain situations, when other factors gain such weight and prominence that they completely overshadow even the latter significance of time. This is why, the text of I-Ching uses everywhere the notion of 'time' more readily in its ancient form, that is, signifying 'season', or, in other words, a kind of 'epoch' of the year, hence many of the qualities attributed to it come out of this rather 'seasonal/agricultural' interpretation of the term. Hence, it is often said, for example: "The seasons are wrong" – so that the active/superior man/woman accepts them as a model for his/her own conduct and actions, where he/she derives two of the most fundamental characteristics of the eternal/perpetual "heaven's and earth's struggle between each other" – and deduces that the first one of these characteristics is incessant change, whereas the second one is the consequent (immutable) relationships this incessant change creates … In effect, such an interpretation leaves some definite room for considering time from the probabilistic standpoint.

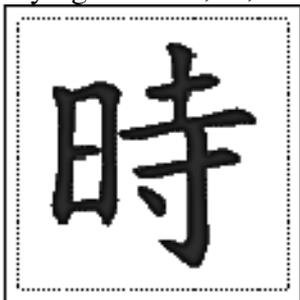

With the above interpretational scheme in mind, conceiving *any* intensive variable as the "odds in favor of the progress" should in effect represent the general approach by G. A. Linhart, the power of which he could also demonstrate by deriving his very simple and universal formula for the heat capacity *vs.* temperature, which is clearly outperforming even the conventional Debye's theory (for the thorough discussion of those G. A. Linhart's works see, for example, [32]). Such a generality claim of the G. A. Linhart's approach can further be supported by substituting the time $t$ in Eqs 2 – 7 by some other intensive variable – for example, by the concentration $c$. With this in mind and following the G. A. Linhart's line of thought, we can formally-mathematically derive the famous empirical equation ingeniously guessed by A. V. Hill [33] more than 100 years ago – and widely used in biochemistry, as well as in pharmacy till now, still causing intensive discussions (see, for example, [34-36]) – to try rationalizing the physical-chemical sense of the Hill equation coefficients this way.

Indeed, we can successfully use the idea by G. A. Linhart to describe the process as the 'dialectic progress-hindrance interplay', when some ligand molecules are binding to some proper sites of a macromolecule. We are herewith trying to describe the fraction of the macromolecule's binding sites saturated by ligand as a function of the ligand concentration. Then, we employ Eqs 2 – 4, with the only intensive variable now being not the time, but $c$ – the free (unbound) ligand concentration, so that the magnitude $S$ in Eq. 3 is now "concentration-dynamic entropy" describing all the possible hindrances to the ligand binding process, whereas $K$ in Eq 4 can be seen in this case as the "coefficient of ligand binding efficiency". With this in mind, the ratio $G/G_i$, is in effect nothing else than the fraction of the macromolecular sites occupied by the ligand molecules, where the ligand can bind to the active site of the receptor macromolecule. Therefore, following Eqs 4 – 6 with this in mind, we arrive at Eq 7, which is now nothing else than just the genuine A. V. Hill [33] equation. We note only, that in this form of Eq 7 $t$ corresponds but to the free (unbound) ligand concentration $c$, whereas $t_S$ is nothing else than the conventional magnitude of the $K_d = (K_A)^n, n \equiv K$, i. e., the ligand concentration producing half-occupation (ligand concentration occupying half of the binding sites), that is, also the microscopic dissociation constant. So that, the magnitude of $K$ in Eq 4 stands now for nothing else than $n$, that is, the genuine "Hill coefficient", describing cooperativity (or possibly other biochemical properties, depending on the context in which the Hill equation is being used).

**The ideas of Georg(e) Augustus Linhart as a specific development of Gilbert Newton Lewis' lifework**

It is of considerable interest to reveal the interconnection between the ideas of Gilbert Newton Lewis and George Augustus Linhart. Gilbert Newton Lewis is very well known for his reformulation of chemical thermodynamics in the mathematically rigorous, but nevertheless readily understandable language (although G. N. Lewis' candidacy was many times nominated for the Nobel prize award, he could never receive it). In his book [37] G. N. Lewis had thoroughly analyzed the true meanings of the 1LT and the 2LT, as well as those of the thermodynamic equilibrium, energy and entropy.

Indeed, in the G. N. Lewis' book chapter entitled "<u>The Power and The Limitations of Thermodynamics</u>" we read:

„*Our book may be introduced by the very words used by Le Chatelier a generation ago:* „<u>*These investigations of a rather theoretical sort are capable of much more immediate practical application than one would be inclined to believe. Indeed the phenomena of chemical equilibrium play a capital role in all operations of industrial chemistry*</u>". He continues: „*Unfortunately there has been such an abuse of the applications of thermodynamics that it is in discredit among the experimentors.*" *If this was true when written by Le Chatelier it is no less true to-day. The widespread prejudice against any practical use of thermodynamics in chemistry is not without reason, for the propagandists of modern physical chemistry have at times shown more zeal than scientific caution.*"

As to the limitations of thermodynamics, G. N. Lewis had communicated the following thoughts: „*The thermodynamics tells us the minimum amount of work necessary for a certain process, but the amount which will actually be used will depend upon many circumstances. Likewise thermodynamics shows us whether a certain reaction may proceed, and what maximum yield may be obtained, but gives no information as to the time required.*" And this is just the logical point where G. A. Linhart had started to work out his „chronodynamical addition" to the conventional thermodynamical theory.

As for the thermodynamic equilibrium, G. N. Lewis had written: „*As a science grows more exact it becomes possible to employ more extensively the accurate and concise methods and notation of mathematics. At the same time it becomes desirable, and indeed necessary, to use words in a more precise sense. For example if we are to speak, in the course of this work, of a pure substance, or of a homogeneous substance, these words must convey as nearly as possible the same meaning to writer and to reader. Unfortunately it is seldom*

*possible to satisfy this need by means of formal definitions; partly because the most fundamental concepts are the least definable, partly because of the inadequacy of language itself, but more particularly because we often wish to distinguish between things which differ rather in degree than in kind. Frequently therefore our definitions serve to divide for our convenience a continuous field into more or less arbitrary regions, — as a map of Europe shows roughly the main ethnographic and cultural divisions, although the actual boundaries are often determined by chance or by political expediency. The distinction between a solid and a liquid is a useful one, but no one would attempt to fix the exact temperature at which sealing-wax or glass passes from the solid to the liquid state. Any attempt to make the distinction precise, makes it the more arbitrary."*

Then, G. N. Lewis was trying to define the notion of the thermodynamic equilibrium and had noted, as follows: „*If it were possible to know all the details of the internal constitution of a system, in other words, if it were possible to find the distribution, the arrangement, and the modes of motion of all the ultimate particles of which it is composed, this great body of information would serve to define what may be called the microscopic state of the system, and this microscopic state would determine in all minutiae the properties of the system. We possess no such knowledge, and in thermodynamic considerations we adopt the converse method. The state of a system (macroscopic state) is determined by its properties, just in so far as these properties can be investigated directly or indirectly by experiment. We may therefore regard the state of a substance as adequately described when all its properties, which are of interest in a thermodynamic treatment, are fixed with a definiteness commensurate with the accuracy of our experimental methods. Let us quote from Gibbs in this connection:* <u>"So when gases of different kinds are mixed, if we ask what changes in external bodies are necessary to bring the system to its original state, we do not mean a state in which each particle shall occupy more or less exactly the same position as at some previous epoch, but only a state which shall be undistinguishable from the previous one in its sensible properties. It is to states of systems thus incompletely defined that the problems of thermodynamics relate."</u>

*The properties of a substance describe its present state and do not give a record of its previous history. When we determine the property of hardness in a piece of steel we are not interested in the previous treatment which produced this degree of hardness. If the metal has been subjected to mechanical treatment, the work which has been expended upon it is not a property of the steel, but its final volume is such a property. It is an obvious but highly important corollary of this definition that, when a system is considered in two different states, the difference in volume or in any other property, between the two states, depends solely upon those states themselves, and not upon the manner in which the system may pass from one state to the other.*

*Most of the properties which we measure quantitatively may be divided into two classes. If we consider two identical systems, let us say two kilogram weights of brass, or two exactly similar balloons of hydrogen, the volume, or the internal energy, or the mass of the two is double that of each one. Such properties are called extensive. On the other hand, the temperature of the two identical objects is the same as that of either one, and this is also true of the pressure and the density. Properties of this type are called intensive. They are often derived from the extensive properties; thus, while mass and volume are both extensive, the density, which is mass per unit volume, and the specific volume, which is volume per unit mass, are intensive properties. These intensive properties are the ones which describe the specific characteristics of a substance in a given state, for they are independent of the amount*

*of substance considered. Indeed, in common usage it is only these intensive properties which are meant when the properties of a substance are being described.*"

This line of thoughts had led G. N. Lewis to the following definition of the termodynamic equilibrium: „*When a system is in such a state that after any slight temporary disturbance of external conditions it returns rapidly or slowly to the initial state, this state is said to be one of equilibrium. A state of equilibrium is a state of rest.*" Again, there clearly remains some indeterminacy with „***rapidly or slowly***". And here is apparently just the logical point, from where the G. A. Linhart's line of thoughts started. Indeed, G. A. Linhart suggested to treat the time just as one of the intensive thermodynamical variables, like, for example, the temperature. Whereas, G. N. Lewis had come similarly to the definition of a ***Partial Equilibrium, Degrees of Stability***: „*Of the various possible processes which may occur within a system, some may take place with extreme slowness, others with great rapidity. Hence we may speak of equilibrium with respect to the latter processes before the system has reached equilibrium with respect to all the possible processes.*" Then, G. N. Lewis had underlined the notion of the „***Equilibrium as a Macroscopic State.***" Specifically, he noted what he means as follows: "*Even here it is desirable to emphasize that by a state of rest, or equilibrium, we mean a state in which the properties of a system, as experimentally measured, would suffer no further observable change even after the lapse of an indefinite period of time. It is not intimated that the individual particles are unchanging.*" But still, there is another very interesting fragment in the book of G. N. Lewis, namely in the chapter devoted to the thermodynamical equilibrium. Specifically, G. N. Lewis stated: „*In practice we often assume the existence of several such equilibrium states toward which a system may tend, all these states being stable, but representing higher or lower degrees of stability. From a theoretical standpoint it might be doubted whether there is any condition of real equilibrium, with respect to every conceivable process, except the one which represents the most stable state. This, however, is not a question which need concern us greatly, nor is it one which we could discuss adequately at this point, without largely anticipating what we shall later have to say regarding the statistical view of thermodynamics.*" Apparently, here is just what had led G. A. Linhart to his implicit idea of a „**fuzzy equilibrium**" (in the sense of interplay between probabilities and possibilities – although the work of Zade [38] was not published that time as yet !), that is, some degree of equilibrium, instead of the conventional (even till nowadays !) ‚crisp' „equilibrium/non-equilibrium" classification ...

Concerning the First Law of Thermodynamics (1LT), G. N. Lewis had expressed the following ideas: „*So, as science has progressed, it has been necessary to invent other forms of energy, and indeed an unfriendly critic might claim, with some reason, that the law of conservation of energy is true because we make it true, by assuming the existence of forms of energy for which there is no other justification than the desire to retain energy as a conservative quantity. This is indeed true in a certain sense, as shown by the explanations which have been given for the enormous, and at first sight apparently limitless, energy emitted by radium. But a study of this very case has shown the power and the value of the conservation law in the classification and the comprehension of new phenomena. It should be understood that the law of conservation of energy implies more than the mere statement that energy is a quantity which is constant in amount. It implies that energy may be likened to an indestructible and uncreatable fluid which cannot enter a given system except from or through surrounding systems. In other words, it would not satisfy the conservation law if one system were to lose energy, and another system, at a distance therefrom, were simultaneously to gain energy in the same amount. If a system gains or loses energy, the immediate surroundings must lose or gain energy in the same amount, and energy may be said to flow*

*into or out of the system through its boundaries. The energy contained within a system, or its internal energy, is a property of the system. The increase in such energy when a system changes from state A to state B is independent of the way in which the change is brought about. It is simply the difference between the final and the initial energy. It is, however, of much theoretical interest to note that the great discovery of Einstein embodied in the principle of relativity, shows us that every gain or loss of energy by a system is accompanied by a corresponding and proportional gain or loss in mass, and therefore presumably that the total energy of any system is measured merely by its mass. In other words, mass and energy are different measures of the same thing, expressed in different units; and the law of conservation of energy is but another form of the law of conservation of mass.*" And here is again just the starting logical point for the idea of G. A. Linhart, that any unidirectional progress in any physical system ought to be paid by a change of the one form of energy to its other form.

The change of one form of energy to another one can be caused and measured by changes in the intensive variables and G. N. Lewis stated in this connection as follows: „*There are two intensive properties, pressure and temperature, which play an important role in thermodynamics, since they largely affect, and often completely determine, the state of a system. Pressure is too familiar an idea to require definition; it has the dimensions of force per unit area, and therefore pressure times volume has the dimensions of energy (force times distance). The concept of temperature is a little more subtle. When one system loses energy to another by thermal conduction or by the emission of radiant energy, there is said to be a flow of heat, a thermal flow. The consideration of such cases leads immediately to the concept of temperature, which may be qualitatively denned as follows: if there can be no thermal flow from one body to another, the two bodies are at the same temperature; but if one can lose energy to the other by thermal flow, the temperature of the former is the greater. This establishment of a qualitative temperature scale is obviously more than a definition. It involves a fundamental principle, to which we have already given preliminary expression in discussing equilibrium, but which we are not yet ready to put in a general and final form. For thermal flow, this principle requires that if A can lose energy to B, B cannot lose it to A; if A can lose to B, and B can lose to C, C cannot lose to A. As in our general discussion of equilibrium, it must be understood that we are dealing with net gains or losses in energy. We do not mean that no energy passes from a cold body to a hot, but only that the amount so transferred is always less than that simultaneously transferred from hot to cold. When we have established the qualitative laws of temperature, we still have a wide freedom of choice in fixing the quantitative scale. Indeed, temperature, as ordinarily measured, or its square, or its logarithm, would equally satisfy these qualitative requirements.*" And this is where G. N. Lewis had apparently stopped, but G. A. Linhart had gone even further, in considering also the time as an intensive thermodynamical variable.

Then, G. N. Lewis was discussing the notions of ‚heat' and ‚work' as the thermodynamically valid forms of the energy change: „*There are two terms, "heat" and "work," that have played an important part in the development of thermodynamics, but their use has often brought an element of vagueness into a science which is capable of the greatest precision. For our present purpose we may say that when a system loses energy by radiation or thermal conduction it is giving up heat; and that when it loses energy by other methods, usually by operating against external mechanical forces, it is doing work. According to the law of the conservation of energy, any system in a given condition contains a definite quantity of energy, and when this system undergoes change, any gain or loss in its internal energy is equal to the loss or gain in the energy of surrounding systems. In any physical or chemical process, the increase in energy of a given system is therefore equal to the heat absorbed from*

*the surroundings, less the work done by the system upon the surroundings. The values of heat and work depend upon the way in which the process is carried out, and in general neither is uniquely determined by the initial and final states of the system. However, their difference is determined, so that if either heat or work is fixed by the conditions under which the process occurs, the other is also fixed. Thus, where the work done by the system is the work of expansion against an external pressure, the expansion may be carried out in such a manner that no heat enters or leaves the system.*" After such deliberations, G. N. Lewis went on with discussing the notions of heat content, heat capacity as well as the units of measuring different forms of energy.

The above-sketched way of thoughts had then led G. N. Lewis to the thorough discussion of the Second Law of Thermodynamics (2LT) and the notion of entropy. The 2LT chapter of his book contains the following deep thoughts: "*After the extremely practical considerations in the preceding chapters, we now turn to a concept of which neither the practical significance nor the theoretical import can be fully comprehended without a brief excursion into the fundamental philosophy of science. Clausius summed up the findings of thermodynamics in the statement, "die Energie der Welt ist konstant; die Entropie der Welt strebt einem Maximum zu", and it was this quotation which headed the great memoir of Gibbs on "The Equilibrium of Heterogeneous Substances". What is this entropy, which such masters have placed in a position of coordinate importance with energy, but which has proved a bugbear to so many a student of thermodynamics ? The first law of thermodynamics, or the law of conservation of energy, was universally accepted almost as soon as it was stated; not because the experimental evidence in its favor was at that time overwhelming, but rather because it appeared reasonable, and in accord with human intuition. The concept of the permanence of things is one which is possessed by all. It has even been extended from the material to the spiritual world. The idea that, even if objects are destroyed, their substance is in some way preserved, has been handed down to us by the ancients, and in modern science the utility of such a mode of thought has been fully appreciated. The recognition of the conservation of carbon permits us to follow, at least in thought, the course of this element when coal is burned and the resulting carbon dioxide is absorbed by living plants, whence the carbon passes through an unending series of complex transformations.*"

After such a remark, G. N. Lewis had thoroughly discussed the philosophical problems connected with the 2LT: "*The second law of thermodynamics, which is known also as the law of the dissipation or degradation of energy, or the law of the increase of entropy, was developed almost simultaneously with the first law through the fundamental work of Carnot, Clausius and Kelvin. But it met with a different fate, for it seemed in no' recognizable way to accord with existing thought and prejudice. The various laws of conservation had been foreshadowed long before their acceptance into the body of scientific thought. The second law came as a new thing, alien to traditional thought, with far-reaching implications in general cosmology. Because the second law seemed alien to the intuition, and even abhorrent to the philosophy of the times, many attempts were made to find exceptions to this law, and thus to disprove its universal validity. But such attempts have served rather to convince the incredulous, and to establish the second law of thermodynamics as one of the foundations of modern science. In this process we have become reconciled to its philosophical implications, or have learned to interpret them to our satisfaction; we have learned its limitations, or better we have learned to state the law in such a form that these limitations appear no longer to exist; and especially we have learned its correlation with other familiar concepts, so that now it no longer stands as a thing apart, but rather as a natural consequence of long familiar ideas.*"

And then, G. N. Lewis had shown one of the possible ways to solve all the philosophical discrepancies introduced by the 2LT, which is till nowadays remaining the common way of thoughts. Specifically, he introduces the **"Preliminary Statement of the Second Law"**, by defining **"The Actual or Irreversible Process"**. Hence, "*The second law of thermodynamics may be stated in a great variety of ways. We shall reserve until later our attempt to offer a statement of this law which is free from every limitation, and shall confine ourselves for the present to a discussion of the law sufficient to display its character and content. Indeed in an early chapter we have already announced the essential feature of the second law when we stated that every system left to itself changes, rapidly or slowly, in such a way as to approach a definite final state of rest. This state of rest (defined in a statistical way) we also called the state of equilibrium. Now since it is a universal postulate of all natural science that a system, under given circumstances, will behave in one and only one way, it is a corollary that no system, except through the influence of external agencies, will change in the opposite direction, that is, away from the state of equilibrium.*" And here, one large problem can be seen immediately, namely, the really fruitful idea of the 'partial equilibrium' introduced by G. N. Lewis, as we have already discussed earlier – and G. A. Linhart had used for some further theoretical developments – didn't find any of its possible applications. But, nevertheless, this is in full accord with all the further developments of thermodynamics (all the well-known theories of irreversible/non-equilibrium thermodynamics – and so on, so forth) …

In accordance with this, G. N. Lewis continued to further develop the above-mentioned logically incomplete representation in his book. Specifically, he stated: "*Before proceeding to a more exact characterization of the second law, let us make sure that there is no misunderstanding of its qualitative significance. When we say that heat naturally passes from a hot to a cold body, we mean that, in the absence of other processes which may complicate, this is the process which inevitably occurs. It is true that by means of a refrigerating machine we may further cool a cold body by transferring heat from it to its warmer surroundings, but here we are in the presence of another dissipative process proceeding in the engine itself. If we include the engine within our system, the whole is moving always toward the condition of equilibrium. A system already in thermal equilibrium may develop large differences of temperature through the occurrence of some chemical reaction, but all such phenomena are but eddies in the general uni-direcional flow toward a final state of rest. The essential content of the second law might be given by the statement that when any actual process occurs it is impossible to invent a means of restoring every system concerned to its original condition. Therefore, in a technical sense, any actual process is said to be irreversible.*" Thus, he had come to the notion of **'The Ideal or Reversible Process'**. Indeed, G. N. Lewis continued as follows: "*When we speak of an actual process as being always irreversible we have had in mind a distinction between such a process and an ideal process which, although never occurring in nature, is nevertheless imaginable. Such an ideal process, which we will call reversible, is one in which all friction, electrical resistance, or other such sources of dissipation are eliminated. It is to be regarded as a limit of actually realizable processes. Let us imagine a process so conducted that at every stage an infinitesimal change in the external conditions would cause a reversal in the direction of the process; or, in other words, that every step is characterized by a state of balance. Evidently a system which has undergone such a process can be restored to its initial state without more than infinitesimal changes in external systems. It is in this sense that such an imaginary process is called reversible.*"

Therefore, G. N. Lewis had logically arrived at the task of defining the **'Quantitative Measure of Degradation'**, that is, the **'Quantitative Measure of Irreversibility'**. And he described his solution to this very important problem, thus coming to the notion of entropy, as follows: "*In viewing the reversible process as the limit toward which actual processes may be made to approach indefinitely, it is implied that processes differ from one another in their degree of irreversibility. It is of the utmost importance to establish a quantitative measure of this degree of irreversibility, or this degree of degradation. So far we have not given a name to our measure of the irreversibility of the standard process. The value of heat-to-temperature ratio, when this process occurs, we shall call the increase in entropy. Thus, entropy has the same dimensions as heat capacity. Our present definition of entropy will be found identical with the definition originally given by Clausius. We have, however, departed radically from the traditional method of presenting this idea, for we have desired to emphasize the fact that the concept of entropy, as a quantity which is always increasing in all natural phenomena, is based upon our recognition of the uni-directional flow of all systems toward the final state of equilibrium. In the ordinary definition of entropy the attention is focussed upon the reversible process and not upon the irreversible process, the existence of which necessitates the entropy concept. For this reason we have based our definition immediately upon an irreversible process, and shall now employ the reversible process only as a means of comparing the degree of degradation, or the increase in entropy, of two irreversible processes.*"

Keeping in mind the above deliberations, G. N. Lewis had underlined, that the entropy, thus defined, is an **'Extensive Property'**. And he stated in this connection: "*In expressing the entropy change during an irreversible process as the difference between the entropy at the end and the entropy at the beginning, we have implied that entropy is a property, and therefore that the entropy change depends solely upon the initial and final states. Indeed this follows directly from our definition, for by whatever irreversible path we proceed from state A to state B, the minimum degradation of the spring-reservoir-system necessary for the return from state B to state A is the same. It is true that we have not shown how to obtain the absolute entropy value of $S_B$ or $S_A$, but only their difference. In the meantime we shall regard the entropy, like the energy and heat content, as a quantity of which the absolute magnitude is undetermined. Moreover, entropy is an extensive property, for we may consider two systems which are just alike, and each of which undergoes the same infinitesimal irreversible process; evidently the change in the standard spring-reservoir-system necessary for their restoration is twice as great as it would be for one of them alone. Since entropy is extensive, we may regard the entropy of a system as equal to the sum of the entropies of its parts. It is therefore important to ascertain how to determine the localization of entropies in the various parts of a system. Owing to the special properties of the standard spring-reservoir-system which we assumed at the outset, it will be convenient to postulate that in any operation of the spring-reservoir-system the entropy changes occur in the reservoir alone, so that if the standard reservoir gains heat from any source by the amount q at the temperature T, the reservoir changes in entropy by the ratio of q/T.*" But then G. N. Lewis had concluded: "*We have seen that the total entropy change in a reversible process is zero. It follows that in such a process the entropy change in any system must be equal and opposite in sign to the entropy change in all other systems involved. In order to study this case further, let us consider the energy changes which occur in a reversible process between some system and the standard spring-reservoir. For the sake of simplicity we shall choose an infinitesimal process. We may sum up our quantitative conclusions regarding entropy. In any irreversible process the total entropy of all systems concerned is increased. In a reversible process the total increase in entropy of all systems is zero, while the increase in the entropy of any*

*individual system, or part of a system, is equal to the heat which it absorbs divided by its absolute temperature. It is important to see clearly that the idea of entropy.is necessitated by the existence of irreversible processes; it is only for the purpose of convenient measurement of entropy changes that we have discussed reversible processes here.*" After presenting all the above thoughts, G. N. Lewis continued, nevertheless, with the reflections about the interconnection between the entropy and probability, for he was apparently feeling the logical deficiencies of the above-sketched 2LT interpetation … Specifically, he wrote: "*The second law of thermodynamics is not only a principle of wide-reaching scope and application, but also it is one which has never failed to satisfy the severest test of experiment. The numerous quantitative relations derived from this law have been subjected to more and more accurate experimental investigation without detection of the slightest inaccuracy. Nevertheless, if we submit the second law to a rigorous logical test, we are forced to admit that, as it is ordinarily stated, it cannot be universally true. It was Maxwell who first showed the consequences of admitting the possible existence of a being who could observe and discriminate between the individual molecules. This creature, usually known as Maxwell's demon, was supposed to stand at the gateway between two enclosures containing the same gas at the same original temperature. If now he were able, by openings and shuttings the gate at will, to permit only rapidly moving molecules to enter one enclosure and only slowly moving molecules to enter the other, the result would ultimately be that the temperature would increase in one enclosure and would decrease in the other. Or, again, we could assume the enclosures filled with air, and the demon operating the gate to permit only oxygen molecules to pass in one direction and only nitrogen molecules in the other, so that ultimately the oxygen and nitrogen would be completely separated. Each of these changes is in a direction opposite to that in which a change normally occurs, and each is therefore associated with a diminution in entropy. Of course even in this hypothetical case one might maintain the law of entropy increase by asserting an increase of entropy within the demon, more than sufficient to compensate for the decrease in question. Before conceding this point it might be well to know something more of the demon's metabolism. Indeed a suggestion of Helmholtz raises a serious scientific question of this character. He inquires whether micro-organisms may not possess the faculty of choice which characterizes the hypothetical demon of Maxwell. If so, it is conceivable that systems might be found in which these micro-organisms would produce chemical reactions where the entropy of the whole system, including the substances of the organisms themselves, would diminish. Such systems have not as yet been discovered, but it would be dogmatic to assert that they do not exist. While in Maxwell's time it seemed necessary to ascribe demoniacal powers to a being capable of observing molecular motions, we now recognize that the Brownian movement, which is readily observable under the microscope, is in reality thermal motion of large molecules. It would therefore seem possible, by an extraordinarily delicate mechanism in the hands of a careful experimentor, to obtain minute departures from the second law, as ordinarily stated. But here also we should depend upon a conscious choice exercised by the experimentor. It would carry us altogether too far from our subject to take part in the long-continued debate on the subject of vitalism; the vitalists holding that there are certain properties of living matter which are not possessed at all by inanimate things, or, in other words, that there is a difference in kind between the animate and the inanimate. However, we may point out that in the lasj^ analysis differences of kind are often reduced to differences in degree. There certainly can be no question as to the great difference in trend which exists between the living organism, and matter devoid of life. The trend of ordinary systems is toward simplification, toward a certain monotony of form and substance; while living organisms are characterized by continued differentiation, by the evolution of greater and greater complexity of physical and chemical structure. In the brilliant investigation of Pasteur on asymmetric or optically active substances, it was shown that a system of optically*

*inactive ingredients never develops optically active sub stances except through the agency of living organisms, or as the result of the conscious choice of an experimentor.*" And then G. N. Lewis concluded: "*Sometimes when a phenomenon is so complex as to elude direct analysis, whether it concern the life and death of a human being, or the toss of a coin, it is possible to apply methods which are called statistical. Thus tables and formulae have been developed for predicting human mortality and for predicting the results of various games of chance, and such methods are applied with the highest degree of success. It is true that in a given community the "expectation of life" may be largely and permanently increased by sanitary improvements, but if a great many individual cases be taken promiscuously from different localities at different times, the mean duration of life, or the average deviation from this mean, becomes more and more nearly constant the greater the number of cases so chosen. Likewise it is conceivable that a person might become so expert in tossing a coin as to bring heads or tails at will, but if we eliminate the possibility of conscious choice on the part of the player, the ratio of heads to tails approaches a constant value as the number of throws increases. The distinction between the energy of ordered motion and the energy of unordered motion is precisely the distinction which we have already attempted to make between energy classified as work and energy classified as heat. Our present view of the relation between entropy and probability we owe largely to the work of Boltzmann, who, however, himself ascribed the fundamental idea to Gibbs, quoting, "<u>The impossibility of an uncompensated decrease of entropy seems to be reduced to an improbability</u>." It would carry us too far if we should attempt to analyze more fully this idea that the increase in the entropy of a system through processes of degradation merely means a constant change to states of higher and higher probability.*"

In analyzing the course of thoughts of Maxwell, Gibbs and Boltzmann in this direction, G. N. Lewis had still come back to his own reflections about the role of irreversible processes in the thermodynamical theory. He continued as follows: "*The mere recognition that such a relationship exists suffices to give a new and larger conception of the meaning of an irreversible process and the significance of the second law of thermodynamics. If we regard every irreversible process as one in which the system is seeking a condition of higher probability, we cannot say that it is inevitable that the system will pass from a certain state to a certain other state. If the system is one involving a few molecules, we can only assert that on the average certain things will happen. But as we consider systems containing more and more molecules we come nearer and nearer to complete certainty that a system left to itself will approach a condition of unit probability with respect to the various processes which are possible in that system. This final condition is the one which we know as equilibrium. In other words, the system approaches a thermodynamic or macroscopic state, which represents a great group of microscopic states that are not experimentally distinguishable from one another. With an infinite number of molecules, or with any number of molecules taken at an infinite number of different times, the probability that the macroscopic state of the system will lie within this group is infinitely greater than the probability that it will lie outside of that group. Leaving out of consideration systems, if such there be, which possess that element of selection or choice that may be a characteristic of animate things, we are now in a position to state the second law of thermodynamics in its most general form: Every system which is left to itself will, on the average, change toward a condition of maximum probability. This law, which is true for average changes in any system, is also true for any changes in a system of many molecules. We have thought it advisable to present in an elementary way the ideas touched upon, in order to give a more vivid picture of the nature of an irreversible process and a deeper insight into the meaning of entropy. It is true we shall not, henceforth, make formal use of the relation between entropy and probability; nevertheless, we shall always*

*tacitly assume that we are dealing with statistical ideas. For example, when calculating solubilities or vapor pressures,*" Therefore, G. N. Lewis was apparently aware of that the problem to give some really valid definitions of the 2LT and entropy had not been satisfactorily solved as yet. Well, and this seems to be just the logical point where G. A. Linhart had started his great work.

Hence, to my mind, the interpretation suggested by G. A. Linhart is in fact very strong, for he used the dialectical viewpoint on every process, that is, on the 2LT itself, because he considered the energy change as the measure of **'progress'**, whereas the entropy as the measure of **'hindrance'** … Along with this, Linhart's standpoint could include the formal mathematical proof of the interconnection between the entropy and probability (as ingeniously guessed by Maxwell, Boltzmann and Gibbs), clearly avoiding all the principal theoretical complications connected with combining the notions of reversibility and irreversibility, as well as the apparent logical difficulties connected with the notion of the thermodynamic equilibrium, as was surely recognized (but still not carefully thought over !) by G. N. Lewis himself.

**The relationship between the ideas of G. A. Linhart and other viable solutions to the 2LT problem**

The 2LT problem was attracting and still attracts attention of many scientists in the field.

And here is how the problem is widely treated till nowadays: "*The development of material world toward complexity and increasing natural diversity violates the second law of thermodynamics and makes it necessary to investigate non-equilibrium processes that may give rise to orderlyness. Now these problems are considered by synergetics*" [39]. Hermann Haken, one of the founders of the synergetics, had expressed it as follows: "*In physics, there is a notion of "concerted effects"; however, it is applied mainly to the systems in thermal equilibrium. I felt that I should introduce a term for consistency in the systems far from thermal equilibrium… I wished to emphasize the need for a new discipline that will describe these processes… Thus, synergetics can be considered as a science dealing with the phenomenon of self-organization*" [40]. This is just nothing else than the usual widespread line of thoughts, which tends to absolutize the 'crisp' notion of thermodynamical equilibrium (that is, either one has a "strict equilibrium", or a "strict non-equilibrium"), just in apparent contrast to the G. N. Lewis' and G. A. Linhart's intuitive idea of 'fuzzy equilibrium' (that is, one has never 'crisp' differences of the equilibrium/non-equilibrium kind, but always some "partial degrees of equilibrium").

The work [39] suggests to describe different mechanisms of the material world self-organization using the concept of homeostatic determinate systems. We shall give here a short description of this very interesting concept, following the work [39].

*The concept of homeostatic determinate systems is fundamental for synergetics. It provides deep insight into physical meaning and genesis of the hierarchy of instabilities in self-organizing homeostatic systems, into the nature of interrelations between instabilities and order parameters.*

*Under homeostatic determinate systems we understand such systems, where the eventual (actual) result of an action is predicted (determined) via the interaction of signals specific for the given system with its memory elements. Structurally such systems include*

*determinate synthesis, choice of an adequate action program, its outcome, and feedback closed on the results-of-action acceptor.*

*The most complex element of any homeostatic determinate system is the determinate synthesis. Here the dominant motivation, conditional (environmental) and causal (triggering) afferentation interact with the memory elements of the system. This results in decision making and choosing a termodinamically determined action program. The information on the result thus achieved goes to the results-of-action acceptor (feedback), where the anticipated and actual results are compared. If the goal is not reached, i.e., the actual result does not fit the anticipated one, homeostatic determinate systems switch over to another programs due to changes in the decision made.*

*This mechanism of the material world self-organization radically differs from the self-organization in dissipative nonequilibrium systems, where dissipative structures emerge due to reinforcement of the corresponding fluctuations. The appearance of such structures was considered by Ilya Prigogine as "orderliness through fluctuations" [41]. Such a mechanism of the self-organization is abundant in the nature, but it cannot underlie the progressive evolution of material world, which is a stable and irreversible process. The nature has chosen the way of self-organization via the formation of homeostatic determinate systems.*

*In the physical determinate systems, the system-forming factors are represented by the well-known four types of interaction:*

- *strong, or nuclear interaction (force) acts only between hadrons that form the atom nucleus (proton, neutron), and does not depend on electric charge of the interacting particles;*
- *weak interaction (force) acts between leptons (electron), between hadrons, between leptons and hadrons, and does not depend on the charge of interacting particles;*
- *electromagnetic interaction (force) acts between electrically charged particles. With the unlike charges it shows up as attraction, whereas with the like charges – as repulsion;*
- *gravitational interaction (force) acts between all the particles without exception and always shows up as attraction. Since elementary particles have small masses, the role of gravitational interaction between them is negligible.*

*Due to the above-mentioned interactions, atoms of all chemical elements exist as systems, which are so stable under natural conditions, that we can investigate their properties and use them. It can be stated that each chemical element (atom) is a unique determinate system with unique properties. The forces serve as memory elements in these systems and determine their stability.*

*To provide stability of the system, feedback should be introduced. The feedback will close between the actual and predicted results of action (ARA and PRA), which are the stable states of the exchange pairs implementing the above-mentioned four types of interaction. The interaction energy in the pairs should tend to the minimum. In determinate systems the interactions can nence function as the results-of-action acceptor.*

*If no stable interaction is observed in the exchange pairs, the action program is corrected. Therewith, spontaneous transitions may occur in the pairs; however, they may occur also with virtual pairs, i. e., the particles-anti-particles pairs borrowed from the*

*vacuum. It is therefore possible to the use the principle of determinism along with the relativistic principle underlying the behavior of elementary particles.*

*Functional pairs in atoms serve as the results-of-action acceptors; therewith, the formation of a stable state is determined by the interaction of atom with the environment (vacuum), involving the virtual particles. Hence, the environment (physical vacuum) ought to be most essential for the stable existence of any atoms and any of the related quantum transformations. Therefore, atoms might be considered as open systems, like, for example, living objects, which continuously exchange mass, energy and information with the environment, i. e., with the quantum vacuum.*

*The concept of determinate systems*, as it is sketched above, *is a new instrument for the analysis of functional states of atoms and a new conceptual apparatus (language) for their description. This way, it is in principle possible to consider statistical (stochastic) processes in quantum physics as determinate, i. e., as the processes based upon specific causal connections and quite predictable results* [39].

Albert Einstein had expressed once, back in 1934, the following expectation: "*I still believe in a possibility of constructing such model of reality, i. e., the theory that expresses the objects themselves, but not only the probabilities of their behavior*" [42]. Interestingly, in my view, the Linhart's way of thinking matches this expectation of Einstein much more perfectly than the concept [39] sketched above, because the latter concept doesn't even explicitly consider the role of the time (or of any other intensive thermodynamical variable) in the processes under study … Further on, the true role of probability theory was questioned by De Finetti already long time ago [43], whereas G. A. Linhart had already used the ideas akin to those by De Finetti well before the latter ones were published at all.

There is also another most recent work [44], where a review of the irreversibility problem in modern physics with new researches is given. Some characteristics of the Markov chains are specified and the important property of monotonicity of a probability is formulated. Then, the behavior of relative entropy in the classical case is considered. Further, the irreversibility phenomena in quantum problems are studied. This work has also paid no special attention to the explicit role of the time (or of any other intensive thermodynamical variable) in the processes under study …

Interestingly, one more review paper [45] has been published to present some of the recent contributions that show the use of thermodynamics to describe biological systems and their evolution, illustrating the agreement that this theory presents with the field of evolution. Organic systems are described as thermodynamic systems, where entropy is produced by the irreversible processes, considering as an established fact that this entropy is eliminated through their frontiers to preserve life. The necessary and sufficient conditions to describe the evolution of life in the negentropy principle are established. Underlining the fact that the necessary condition requires formulation, which is founded on the principle of minimum entropy production for open systems operating near or far from equilibrium, other formulations are mentioned, particularly the information theory, the energy intensiveness hypothesis and the theory of open systems far from equilibrium. Finally, suggesting the possibility of considering the lineal formulation as a viable alternative; that is, given the internal constrictions under which a biological system operates, it is possible that the validity of its application is broader than it has been suggested. But, again, there is absolutely no discussion of the implicit role of the time time (or of any other intensive thermodynamical

variable) in the processes under study …

Finally, a very interesting paper has been published most recently [46], where the author studies the universal efficiency at optimal work with the help of the Bayesian statistics and finds that, if the work per cycle of a quantum heat engine is averaged over an appropriate prior distribution for an external parameter, the work becomes optimal at Curzon-Ahlborn (CA) efficiency. More general priors yield optimal work at an efficiency which stays close to (CA) value, in particular near equilibrium the efficiency scales as one-half of the Carnot value. This feature is analogous to the one recently observed in literature for certain models of finite-time thermodynamics. Further, the use of the Bayes' theorem implies that the work estimated with posterior probabilities also bears close analogy with the classical formula. These findings suggest that the notion of prior information can be used to reveal thermodynamic features in quantum systems, thus pointing to a connection between thermodynamic behavior and the concept of information. I have entered an intensive discussion with the author of this paper and tried to persuade him, that his approach is in fact very much akin to the Linhart's way of thoughts …

**Conclusions**

George Augustus Linhart was definitely able to successfully work out the true foundations of thermodynamics and could thus outdistance many famous thermodynamicists of his time and even the later ones … Linhart's view of the Second Law of Thermodynamics was and is extremely fruitful. Using the Linhart's line of thoughts it is possible to formally derive the mathematical expression for the famous Boltzmann's entropy, for the heat capacity (which outperforms the famous Debye's formula), for the well-known equation describing the ligand binding to macromolecules, that is, the equation just guessed by Archibald Hill long time ago and never formally-mathematically derived by anybody … The Linhart's point of view enables us to treat the time in rather simple and natural terms of thermodynamics, just as one of the numerous intensive variables, without reverting to the "Arrow of Time" and all the problems of the "Ergodicity". Following G. A. Linhart's ideas enables us to treat any of the possible intensive variables on the same lines, and hence, his approach is definitely of general significance – at least for the thermodynamics as a whole.